\documentclass[prb,twocolumn]{revtex4}
\usepackage{graphicx}
\usepackage{subfig}
\usepackage{epsfig}
\usepackage{amsmath}
\usepackage{amsfonts}
\usepackage{amssymb}
\usepackage[version=3]{mhchem}
\usepackage{siunitx}
\usepackage{hyphenat}
\usepackage{microtype,amsmath}
\usepackage{ragged2e}

\providecommand{\U}[1]{\protect\rule{.1in}{.1in}}

\graphicspath{ {plots/} }

\begin{document}
\sloppy

\title{Pair creation as a source of longitudinal chiral magnetoconductivity}

\author{J. L. Acosta Avalo$^{*}$}
\author{S. Montesino Castillo$^{*\dagger}$}
\author{E. E. Garc\'{\i}a Reynaldo$^{**}$}

\affiliation{%
$^{*}$\textit{Instituto Superior de Tecnolog\'{\i}as y Ciencias Aplicadas. Universidad de la Habana, Ave Salvador Allende, No. 1110, Vedado, La Habana, 10400 Cuba}\\
$^{**}$\textit{Instituto de Cibern\'{e}tica, Matem\'{a}tica y F\'{\i}sica, Calle E esq 15, No. 309, Vedado, La Habana, 10400 Cuba}\\
$^{\dagger}$\textit{Present address: Institute for Theoretical and Mathematical Physics, Lomonosov Moscow State University, Moscow, Russia}
}

\begin{abstract}

\small{We demonstrate that chiral transport in a strongly magnetized electron-positron plasma can arise dynamically from dissipative pair-creation processes encoded in the imaginary part of the photon polarization tensor within one-loop finite-temperature quantum electrodynamics (QED). In the kinematic region corresponding to longitudinal photon absorption, real electron-positron pair production induces axial charge nonconservation and generates an electric current parallel to the magnetic field, without requiring the introduction of an external chiral chemical potential. This provides a microscopic mechanism for chiral magnetic transport, offering an alternative to hydrodynamic or anomaly-based effective descriptions in which chirality imbalance is typically introduced as an external input. We derive an explicit expression for the longitudinal magnetoconductivity associated with this process and show that it exhibits an approximately quadratic dependence on the magnetic field only within a restricted intermediate regime. This behavior emerges from the dominance of the lowest
Landau levels as a  characteristic of negative longitudinal magnetoresistance. We further analyze how Pauli blocking regulates the pair-creation phase-space and demonstrate that the dynamically generated chiral imbalance is suppressed at high frequencies, revealing a transition between chiral-active and non-chiral-active regimes. Our results connect microscopic QED processes with anomaly-related transport phenomena in strongly magnetized relativistic plasmas, where pair creation provides a dynamical source for chiral imbalance.}

\bigskip
\end{abstract}

\maketitle

\section{INTRODUCTION}
\label{sec1}

\ \

The influence of \textit{magnetic fields} on relativistic quantum systems is fundamental, as such fields profoundly alter particle dynamics, symmetry properties, and transport phenomena \cite{Gusynin1996,Shovkovy2013}. In neutron stars, ultra-strong magnetic fields reaching $10^{14}$--$10^{15}\,\mathrm{G}$ affect the equation of state and the behavior of charged particles in dense matter \cite{Duncan1992, Lai2001,Harding2006,Turolla2015,Beloborodov2012}. In quasars and astrophysical jets, magnetic fields guide relativistic outflows and influence radiation mechanisms \cite{Blandford1977}. In relativistic heavy-ion collisions, extremely strong and short-lived magnetic fields---estimated to reach up to $10^{18}$--$10^{19}$ G---can strongly affect the dynamics of the quark-gluon plasma, leading to charge separation and modifications in collective flow observables \cite{Kharzeev2008,Skokov2009}. Therefore, understanding quantum fields in magnetized backgrounds has become essential for connecting high-energy theory with observable phenomena in both astrophysical and laboratory environments.

Building upon this foundation, recent developments have drawn increasing attention to the role of strong magnetic fields in systems with chiral fermions\cite{Kharzeev2014, Fukushima2008,Landsteiner2016,Kharzeev2016,Xiong2015,Armitage2018,Nag2020}. These studies have unveiled a rich landscape of anomalous transport phenomena that emerge from the quantum structure of field theory. Among these, the most prominent is the \emph{chiral magnetic effect} (CME)\cite{Fukushima2008}, in which an electric current is induced along a magnetic field  in the presence of a chiral imbalance. This effect has significant implications not only in astrophysical and cosmological contexts but also in heavy-ion collisions, where such imbalances and intense magnetic fields can transiently coexist\cite{Li2026}.

The theoretical underpinning of these phenomena lies in the \emph{chiral anomaly}, where the divergence of the axial current becomes nonzero in the presence of gauge fields with nontrivial topology\cite{Adler1, Bell}. This so-called Adler-Bell-Jackiw anomaly underlies transport effects such as the CME, in which the longitudinal magnetoconductivity scales quadratically with the magnetic field, \(\sigma \propto B^{2}\), a typical manifestation of negative longitudinal magnetoresistance \cite{Kharzeev2016,Xiong2015}.

In many studies of the CME, chirality imbalance is introduced through a nonvanishing chiral chemical potential $\mu_5$. In contrast, our approach does not require an externally imposed chiral asymmetry. Although we introduce an ordinary electromagnetic chemical potential $\mu$, as in Ref.\cite{Vilenkin}, our results remain valid for $\mu=0$. Instead, the chiral imbalance is generated dynamically through electron-positron pair creation induced by the absorption of longitudinal pseudovector photon modes propagating along an external magnetic field. Within this framework, the longitudinal mode is directly associated with the chiral charge density in a magnetized QED medium.

We emphasize that, while magnetohydrodynamics (MHD) successfully describes large-scale dynamics\cite{Son2009}, it does not capture the \emph{microscopic quantum effects} such as the pair creation process. In this sense, our approach, based on a finite-temperature ($T\neq0$, $T$ in energy units) one-loop QED framework, allows us to describe this fundamental process, and provides a connection between this process and measurable transport signatures like \emph{negative longitudinal magnetoresistance}.

Another important point concerns the commonly used approximation of vanishing fermion mass, which is convenient since particles then have definite helicity, but as noted in Ref.\cite{Vilenkin}, this is not realistic.  For instance, it becomes problematic in the presence of magnetic fields, where charged massless particles exhibit divergent magnetic moments. Our formalism describes massive fermions, and the massless limit is interpreted as an effective mass in a  relativistic medium. 

The above discussion  forces us to revisit the notion of chiral symmetry breaking  in massive QED. Unlike Ref.\cite{Vilenkin}, we distinguish between chiral symmetry breaking due to nonzero mass (which we may refer to as scalar or dynamical), compatible with thermodynamic equilibrium (since at each instant in equilibrium an equal number of left and right particles is expected on average), and pseudoscalar chiral symmetry breaking. The latter arises from gauge field configurations with nontrivial topology and is associated with the invariant $\mathfrak{G} = \frac{1}{4} \mathbf{E} \cdot \mathbf{B}$ (systems with parallel electric-$\mathbf{E}$ and magnetic-$\mathbf{B}$ fields), which leads to nonequilibrium processes of electric current and transport of charge. The axial charge refers to the average momentum-space density of particles and antiparticles sharing a common helicity. The quantity $\mathfrak{G}$ acts as an effective source term for the chiral density in the  continuity equation, leading to the breaking of the previously existing statistical chiral balance of the densities of charged particles\cite{Fukushima2008, Kharzeev2014, Acosta2016}.

We consider a uniform field $\mathbf{B}$ and Minkowski spacetime, neglecting curved spacetime and dipole field structures such as those existing in neutron stars. A stationary $e^{+}e^{-}$ plasma, ignoring relativistic outflows, rotation, and turbulent dynamics, is also considered. These assumptions limit its direct application to realistic astrophysical compact objects. However, our work presents a microscopic mechanism, which identifies longitudinal photon absorption as a viable pathway for pair creation and chiral symmetry breaking, leading to a chiral current along $\mathbf{B}$ and a calculable negative longitudinal magnetoresistance. Crucially, we show this mechanism persists in a low-temperature regime near the magnetized quantum vacuum, where the existence of multiple chiral
configurations is discussed. Our results also provide additional microscopic foundations for chiral transport beyond MHD, with possible implications in astrophysical scenarios like relativistic jets and heavy-ion collisions. Therefore, the value lies not in direct phenomenological application, but in identifying microscopic mechanisms that may be incorporated into more realistic astrophysical models through effective transport coefficients or as source terms in extended MHD frameworks.

This paper is organized as follows. Section~\ref{sec:mechanism1} presents the dynamical generation of chiral imbalance from pair creation by longitudinal photons.  Section~\ref{sec:vacuum} discusses the regime near the magnetized QED vacuum. Section.~\ref{sec:conductivity} derives the longitudinal magnetoconductivity and analyzes its magnetic-field dependence, and Section~\ref{sec:conclusions} contains our final remarks.

\section{Dynamical generation of chiral imbalance from pair creation}
\label{sec:mechanism}

We examine a $e^{+}e^{-}$
 plasma at finite temperature immersed in an external constant field
\textbf{B}  parallel to the $x_3$ axis. In this system, electrons and positrons move in bound states characterized\cite{Acosta2015,Acosta2016} by energy levels given by\renewcommand{\thefootnote}{$\dagger$}\footnotemark\footnotetext{We use natural units, $\hbar = c = 1$, except where explicitly stated.}\renewcommand{\thefootnote}{\arabic{footnote}}

\begin{equation}\label{espectro/energia}
\varepsilon_{n_{l},p_{3}}=\sqrt{p^{2}_{3}+m^{2}+|e|B(2n_{l}+1-sgn(e) s_{3})},
 \end{equation}

 \noindent where $p_3$ is the momentum along $\textbf{B}$, $m$ is the electron mass, $s_{3}=\pm 1$ are spin eigenvalues along $x_{3}$ and $n_{l}=0,1,...$ are the Landau quantum numbers. These are two-fold spin degenerate, except the ground state $\varepsilon_{0}$ in which $n_{l}=0$, and for electrons it is $s_{3}=-1$ and for positrons $s_{3}= 1$.  Quantum states are also degenerate with regard to the orbit's center coordinates \cite{John}. Higher Landau quantum numbers contribute both to paramagnetic and diamagnetic terms \cite{prl2000}, but we will make here the fundamental assumption that  $\sqrt{2eB }\gg\mu, T$, which ensures a large Landau-level separation and supports the dominance
of the lowest Landau levels. This condition is independent of the degeneracy parameter $\mu/T$.

In a quantum relativistic $e^{+}e^{-}$  system subjected to a field $\textbf{B}$, the structure of the photon self-energy tensor determines the propagation modes in both the $C$-invariant and non-invariant cases \cite{Shabad1,Hugo2,Hugo3,Shabad}, as well as the electric transport phenomena. For  $\mu\neq0$, a positive ionic background is assumed to ensure overall charge neutrality, although its dynamical effects can be neglected. The presence of $\textbf{B}$ breaks the spatial symmetry, leaving  invariance for rotations around $\bf{B}$, and for space translations along it. This determines different eigenmodes according to the direction of propagation. 

We recall that in a medium, as different from vacuum, there is a nonvanishing four-velocity vector $u_{\mu} \neq 0$. We must recall that also in the absence of external fields in the charged medium there are three electromagnetic modes, two transverse and one longitudinal \cite{Fradkin2}. The two transverse modes correspond to photon spin projections  $\pm 1$ along its momentum $\bf{k}$, their dispersion equations do not lie on the light cone. The  third mode is a pure electrical longitudinal one (zero spin) and its dispersion equation  in the infrared limit has a solution for $\omega=0, \textbf{k}\neq 0$  which accounts for the Debye mass screening \cite{Fradkin2}, having close formal analogy to the Yukawa force \cite{Acosta2016}. The dispersion equations for these modes can be solved in any direction. In particular, for propagation parallel to $\mathbf{B}$---the case of interest here---the chiral effect associated with longitudinal photons was identified in Ref. \cite{Hugo3,Acosta2016}, while for transverse modes the relativistic Hall conductivity and the Faraday effect were derived in Ref.\cite{Hugo6,Lidice}.

We analyze the pair creation process in a magnetized medium using the Schwinger-Dyson equation for the photon in Fourier space, with $\nu=1,2,3,4$,  as our starting point \cite{Fradkin2}

\begin{equation}\label{ecuacion/SD}
  [k^{2} g_{\mu\nu}-\Pi_{\mu\nu}(k|A_{\mu}^{ext})]a^{\nu}(k)=0,
\end{equation}

\noindent where $g_{\mu\nu}$ is the Minkowski metric, in the form $g_{\mu\nu}=(1,1,1,-1)$ (it corresponds to the analytic continuation $k_{4}\rightarrow i\omega$ from Euclidean metric), and $k^{2}=k_{3}^{2}+k_{\bot}^{2}-\omega^{2}$. Here $k_{3}$ and $k_{\bot}$  are respectively the components of the photon four-momentum in directions parallel and perpendicular to $\textbf{B}$, and $\omega$ its energy. The total external electromagnetic field is  $A^{ext}_{\mu}+a_{\mu}$, where $a_{\mu}$ is a small perturbative radiation field (its electric field  $E\ll B$).

The quantum corrections are given by the photon self-energy tensor $\Pi_{\mu\nu}(k|A_{\mu}^{ext})$, which was calculated in magnetized vacuum and in the one-loop approximation in Ref.\cite{Shabad1,Schwinger1951}, and in a magnetized medium in Ref.\cite{Hugo2,Hugo3,Shabad}. According to Ref.\cite{Hugo2,Shabad1}, the diagonalization of the photon self-energy tensor in vacuum and in a medium  leads to the
equation: $\Pi_{\mu\nu} b^{\nu(i)}=\eta_{i}b_{\mu}^{(i)}$, which has three non-vanishing eigenvalues $\eta_{i}$ and three eigenvectors $b_{\mu}^{(i)}$ for $i = 1, 2, 3,$ (the photon four-vector $k_{\nu}$ has zero eigenvalue) corresponding to three photon propagation modes (see (\ref{eigenmodes}) in Appendix \ref{sec8}). For each mode one obtains a dispersion law\cite{Shabad1,Hugo2,Hugo3,Shabad}
$k^{2}=\eta_{i}(k_{3},k_{\bot},\omega,B)$. In addition to the two transverse modes, there is a longitudinally polarized mode\cite{Acosta2015,Acosta2016} along $\textbf{B}$ given by the pseudovector:  $b_{\mu}^{(2)}(k)=a c_{\mu}^{(2)}$  . Here $c_{\mu}^{(2)}=R_{2}(F^{*}k)_{\mu}$ is a normalized pseudovector  (the normalization parameter is $R_{2}=1/Bz_{1}^{1/2}$, see (\ref{vect-orton})), and $F^{*}_{\mu\nu}$ is the dual of the electromagnetic field tensor $F_{\mu\nu}$. The parameter $a$ (which has dimension of vector potential) is determined by the applied perturbative electric field. Its electric polarization  vector\cite{Shabad1} (see (\ref{ecuaciones-campo electrico-magnetico}) in Appendix \ref{sec8}) is along $\textbf{B}$

\begin{equation}\label{campo-electrico}
  \textbf{E}_{B}=E^{(2)}\textbf{e}_{B}=a(k_{3}^{2}-\omega^{2})^{\frac{1}{2}}\textbf{e}_{B},
\end{equation}

\noindent where $\textbf{e}_{B}=\textbf{B}/B$ is a unit pseudovector. The longitudinal mode does not lie on  the light cone\cite{Shabad1}, that is $k^{2}_{3}-\omega^{2}\neq 0$. From now on we will denote $z_{1}=k_{3}^{2}-\omega^{2}$.                                                                                                                                                                           
The electromagnetic current as a function of $A^{ext}_{\mu}+
a_{\mu}$ depends on the two relativistic invariants: $\mathfrak{F}=B^{2}/2$ and $\mathfrak{G} =\textbf{B}\cdot\textbf{E}$. Notice that for the case of propagation along $\textbf{B}$, the pseudoscalar $\mathfrak{G}\neq0$ only for the longitudinal mode $b_{\mu}^{(2)}$, independently of the $C$-symmetry of the system.

An expansion of the electromagnetic current density as a functional series of $a_\nu$ gives:

\begin{align}\label{desarr-corriente}
  j_{\mu}(A^{ext}_{\mu}+
a_{\mu})=j_{\mu}(A^{ext}_{\mu}) + \frac{\delta j_{\mu}}{\delta A_{\nu}^{ext}}a_{\nu}+... \hspace{2mm} ,
\end{align}

\noindent its linear term in $a_{\nu}$ is \cite{Hugo6,Lidice}

\begin{equation}\label{corriente-tensor de conductividad}
  j_{i}=\Pi_{i\nu}a^{\nu}=Y_{ij}E_{j},
\end{equation}

\noindent where $E_{j}=i(\omega a_{j}-k_{j}a_{0}) $ is the electric field, with $a_{4}=ia_{0}$ and $k_{4}=i\omega $; moreover, $j_{\mu}(A^{ext}_{\mu})= N_{0}\delta_{\mu 4}$, with $N_0$ the net density of charged particles in the ground state. The term $Y_{ij}=\Pi_{ij}/i\omega$ is the complex conductivity. The third term in (\ref{corriente-tensor de conductividad}) comes from the second
one by using the four-dimensional transversality of $\Pi_{\mu\nu}$
due to gauge invariance, $\Pi_{\mu\nu}k^{\nu}=0$ \cite{Shabad1,Hugo2}. In (\ref{desarr-corriente}), $a_{\mu}$ is in general a linear function of the eigenmodes $b_{\mu}^{(i)}$. Below we restrict to the case where the eigenvector is
$a_{\mu}=b_{\mu}^{(2)}$, for which the electric field vector is parallel to $\bf{B}$. Notice that only terms contaning an odd number of $b_{\mu}^{(2)}$ legs in (\ref{desarr-corriente}) lead to pseudovector terms.

Now observe that in the linear approximation in $E_{i}$ ( equation (\ref{corriente-tensor de conductividad})), and from the eigenvalue equation of $\Pi_{\mu\nu}$, one also obtain\cite{Acosta2015,Acosta2016}:

 \begin{equation}\label{corriente-escalar s}
   j_{i}=\Pi_{i\nu}a^{\nu}=s b_{i}^{(2)},
 \end{equation}

\noindent where we can write the scalar $s = c^{(2)\mu}\Pi^{\nu}_{\mu}c^{(2)}_{\nu}$, which is the eigenvalue of the photon self-energy tensor corresponding to the longitudinal mode (see (\ref{escalares}) in Appendix \ref{sec8}). The scalar $s$ in the one-loop approximation is
\cite{Shabad1,Hugo2,Hugo3,Shabad,Acosta2016}:

\begin{multline}\label{escalar-s}
\begin{split}
  s=&-\frac{e^{3}B}{\pi^{2}}\,\sum_{n=0}^{\infty}\int_{-\infty}^{\infty}\frac{dp_{3}}
  {\varepsilon_{q}}\,[\alpha_{n}\varepsilon^{2}_{n,0}(2p_{3}k_{3}+z_{1})]\\
 &\times\frac{[n^{p}(\varepsilon_{q})+n^{e}(\varepsilon_{q})-1]}{4z_{1}p_{3}(p_{3}
 +k_{3})+z^{2}_{1}-4 \omega^{2} \varepsilon^{2}_{n,0}}\,,
 \end{split}
\end{multline}

\noindent where  $n^{e,p}=[1 + e^{(\varepsilon_{q} \mp \mu)/T}]^{-1}$ are the electron and positron distribution functions in momentum space,  $\varepsilon_{n,0}=\sqrt{m^{2}+|e|Bn}$, with $n=2n_{l}+1-sgn(e) s_{3}$, $\alpha_{n}=2-\delta_{n,0}$ and $q=(n,p_{3})$. Here the $-1$ inside the square brackets accounts for the quantum vacuum limit $(\mu,T)\rightarrow(0,0)$.

The current $j_{i}$ in (\ref{corriente-escalar s}) is also a pseudovector (since $b_{\nu}^{(2)}$ is a pseudovector), which is a necessary condition for the breaking of chiral symmetry \cite{Acosta2016}.  Charged fermions interacting with the longitudinal mode exchange energy by the transfer of momentum $k_3$, while the Landau quantum numbers remain unchanged \cite{Hugo6}. Then, we may consider the fermion interaction with the longitudinal mode as a problem in $(1+1)$ dimensions, which is strictly valid if we restrict to the lowest Landau level (LLL).

It is easy to find a gauge transformation (in which one  obtains $b_{3}^{(2)}=(k_{4}/z_{1}) E_3$) leading to $j_3=s(k_4/z_1)E_3$ (from (\ref{corriente-escalar s})), with $E_{3}= E^{(2)}(\textbf{e}_{B}\cdot\textbf{e}_{3})$. This equation is equivalent to

\begin{equation}\label{jota3}
j_{3}=\frac{\Pi_{33}}{k_{4}}E_{3},
\end{equation}

\noindent which is deduced from relation $\Pi_{33}=s(k_{4}^{2}/z_{1})$, where the expression $s = c^{(2)\mu}\Pi^{\nu}_{\mu}c^{(2)}_{\nu}$ and the two-dimensional transversality\cite{Acosta2016} $\Pi_{\mu\nu}k^{\nu}=0$ (with $\mu, \nu=3,4$) were used. Taking into account the above, equation (\ref{corriente-escalar s}), and using both the  transversality condition of $\Pi_{\mu\nu}$ and the identity $\gamma^{\mu}\gamma^{5}=-\epsilon^{\mu\nu}\gamma_{\nu}$ valid in two dimensions\cite{Peskin1995}, we can write the following expression\cite{Acosta2016} for the nonconservation of the axial current $j_A^\mu$

\begin{equation}\label{axial anomaly}
  k_\mu j_A^\mu=\frac{z_1}{k_4}j_3\neq0.
\end{equation}

We remark that while $k_\mu j^\mu=0$ expresses the conservation of the vector current, (\ref{jota3}) and (\ref{axial anomaly}) highlight the role of $E_{3}$ in the breaking of the chiral symmetry, without the need for any preexisting electric charge asymmetry. This produces an electric current along $\textbf{B}$ associated with a chiral magnetic transport. From now on we will restrict to  real frequency and momentum ($k_{3}^{2}>z_{1}$).

We consider only the real part of $j_3$, since  we will restrict ourselves to the imaginary part of $\Pi_{\mu\nu}$. Then we can express relation (\ref{axial anomaly})  induced by longitudinal photons  \cite{Acosta2016} as:

\begin{equation}\label{axial anomaly-1}
  k_\mu j_A^\mu=i\rho\, Im[s]\,E^{(2)},
\end{equation}

\noindent where $\rho=(\textbf{e}_{B}\cdot\textbf{e}_{3})$ is a pseudoscalar factor, and the expression for the imaginary part of the scalar $s$  in one-loop approximation\cite{Hugo3} (details are given in Appendix \ref{sec9}) is:

\begin{align}\label{Imag-s}
Im[s]=-\frac{e^{3}B}{2\pi}\sum_{n=0}^{\infty}\frac{\alpha_{n}\,\varepsilon_{n,0}^{2}}{\Lambda}[\theta_1\,\Delta N(\varepsilon_{r})+\theta_2\,\Delta H(\varepsilon_{s})],
\end{align}

\noindent here $\Lambda=\sqrt{z_{1}(z_{1}+4\varepsilon_{n,0}^{2})}$, $\theta_1=\theta(z_1)$ and $\theta_2=\theta(-4\varepsilon_{n,0}^{2}-z_{1})$ are the Heaviside step functions. The term  

\begin{align}\label{excitation-particles}
 	\Delta N=[N(\varepsilon_{r})-N(\varepsilon_{r}+\omega)],
 \end{align}
 
\noindent where  $N(\varepsilon_{r})=n^{e}(\varepsilon_{r})+n^{p}(\varepsilon_{r})$, accounts for the excitation of particles $[\varepsilon (p_3,n)\longrightarrow \varepsilon (p_3+k_3,n)]$ by increasing their momentum along $\bf{B}$ (region $z_1>0$). The quantity

\begin{align}\label{creation-particles}
 	\Delta H(\varepsilon_{s})=[H(\varepsilon_{s})+H(\omega-\varepsilon_{s})-2],
 \end{align}

\noindent with $H(\varepsilon_{s})=n^{e}(\varepsilon_{s})+n^{p}(\omega-\varepsilon_{s})$, represents the net pair-creation phase-space factor (region $z_{1}<-4\varepsilon^{2}_{n,0}$). The $-2$ inside the square brackets accounts for the quantum vacuum limit. Both processes are associated with the interaction with the longitudinal mode, the Landau quantum numbers being unchanged \cite{Hugo6}. Pauli's exclusion principle\cite{Prakapenia2023} demands that vacant states must exist both for the occurrence of excitation and pair creation processes for fixed $n$ and $T$.  The region $-4\varepsilon^{2}_{n,0}<z_{1}<0$ corresponds to a transparency region on the $z_{1}$ axis for real frequencies. The energies $\varepsilon_{s}=(\omega z_{1}+|k_{3}|\Lambda)/2z_{1}$ and
$\varepsilon_{r}=(-\omega z_{1}+|k_{3}|\Lambda)/2z_{1}$ (with $r,s=(n,\omega,k_{3})$) are the
fermion energies in a magnetic field in terms of the longitudinal mode energy $\omega$ and momentum $k_{3}$ (see  (\ref{energia-ecxitacion}) and (\ref{energia-cracion}) in Appendix \ref{sec9}).

 \subsection{Dynamical axial charge production via longitudinal photons}
\label{sec:mechanism1}

We will now discuss how the absorption of longitudinal photons in a magnetized relativistic medium can induce the creation of electron-positron pairs, leading to a chiral imbalance even in the limit of massless fermions. This mechanism could be of particular interest in astrophysical environments where strong magnetic fields and relativistic plasmas coexist, such as in the magnetospheres of neutron stars or in relativistic jets from active galactic nuclei.  While MHD has provided valuable insights into such systems, it does not capture the microscopic dynamics of pair creation. In this sense, our approach allows us to describe this process within a consistent quantum field theoretical framework.

The photon polarization operator  $\Pi_{\mu\nu}(k|A_{\mu}^{ext})$ presented above describes the creation of virtual electron-positron pairs by real photons, followed by annihilation back into  photons with the same quantum numbers. In a hot and dense medium, this process can become real if the photon energy is sufficiently large. Another process is possible, where a longitudinal photon may be absorbed (or emitted) by an electron or positron, which subsequently emits (or absorbs) another photon with parameters identical to the first, and the electron or positron returns to its initial state. This process could occur in an appropriate medium in the region $z_1>0$.

 Here, we are only interested in the region $z_{1}<-4\varepsilon^{2}_{n,0}$, where  absorption by longitudinal photons is due solely to pair creation. Since $|\omega|>|k_3|$, we have  $|z_{1}|>\Lambda$. Then, for $\omega>0$, the energy associated with particle scattering,  $\varepsilon_{r}$, becomes negative, leading to unphysical states, while the energy for pair creation, $\varepsilon_{s}$, is positive   (see  (\ref{energia-ecxitacion}) and (\ref{energia-cracion}) in Appendix \ref{sec9}). From (\ref{axial anomaly-1}), we obtain:

 \begin{align}\label{axial anomaly-pair creation}
 	k_{\mu} j^{\mu}_A = - \frac{i e^{3}}{2\pi} \sum_{n=0}^{\infty}\alpha_n \frac{\varepsilon^{2}_{n,0}\,\Delta H(\varepsilon_{s})}{\sqrt{z_1(z_1+4\varepsilon^{2}_{n,0})}}\hspace{1mm}(\textbf{E}^{(2)}\cdot \textbf{B}).
 \end{align}

 Equation~(\ref{axial anomaly-pair creation}) shows explicitly that axial charge is dynamically generated whenever the pair creation phase space is open. Importantly, chiral imbalance is entirely determined by the absorptive part of the polarization tensor and the pair creation kinematics. No chiral chemical potential $\mu_5$ has been introduced. The imbalance therefore emerges dynamically from real pair creation in the magnetized medium. Unlike the standard chiral magnetic effect current, which is universal and fixed by a preexisting $\mu_5$, the present mechanism depends explicitly on the photon frequency $\omega$, the longitudinal momentum $k_3$, and the thermal occupation factors through $\Delta H$. This demonstrates that the chiral imbalance obtained here has a genuinely dynamical origin tied to pair creation by longitudinal photons rather than to an externally imposed chiral asymmetry.

The analogy between our derived relation (\ref{axial anomaly-pair creation}) and the Adler–Bell–Jackiw anomaly suggests a connection between longitudinal photon absorption and anomaly-related quantum effects. In vacuum QED the divergence of the axial current is governed by the Adler--Bell--Jackiw anomaly. In contrast, the result derived here emerges from the absorptive part of the photon polarization tensor in a magnetized medium. The nonvanishing imaginary part corresponds physically to real electron--positron pair creation by longitudinal photons. Therefore, the axial charge production rate obtained in this work should be interpreted as a medium--induced dynamical effect rather than as the universal vacuum anomaly. The anomaly fixes the allowed chiral structure of the theory, while the magnetized medium and the longitudinal mode provide the microscopic channel through which axial charge is dynamically generated.

 	\begin{figure}[h]
 		\centering
 		\includegraphics[width=8.5cm]{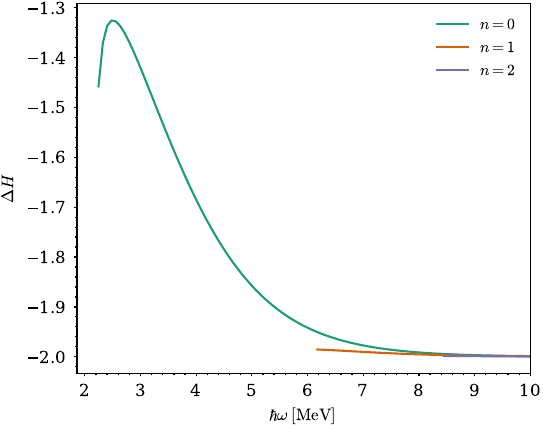}
 \caption{\protect\justifying
Net pair-creation phase-space factor
$\Delta H$
as a function of the longitudinal photon energy $\hbar\omega \leq\, 10\, \mathrm{MeV} < \hbar\omega_{\rm thr}^{(n=3)}$
for Landau levels $n=0-2$.
Results are shown for a longitudinal momentum\cite{Beloborodov2013}
$\hbar ck_3 = 2\,\mathrm{MeV}$,
magnetic field\cite{Harding2006,Turolla2015,Kaspi2017} $B = 7\cdot10^{14}\,\mathrm{G}$,
chemical potential\cite{Svensson1984} $\mu \simeq mc^2$,
and temperature\cite{Svensson1984,Potekhin2015} $T\simeq m c^2$.
The vacuum limit corresponds to $\Delta H=-2$. Photon energies in the MeV range arise naturally in pair cascades and high-energy emission processes in magnetospheres \cite{Beloborodov2013,Daugherty1983,Harding2006,TATAROGLU2025}.
}\label{deltaH/w-diferente-n}
 	\end{figure}
 	
 	\begin{figure}[h]
 		\centering
 		\includegraphics[width=8.5cm]{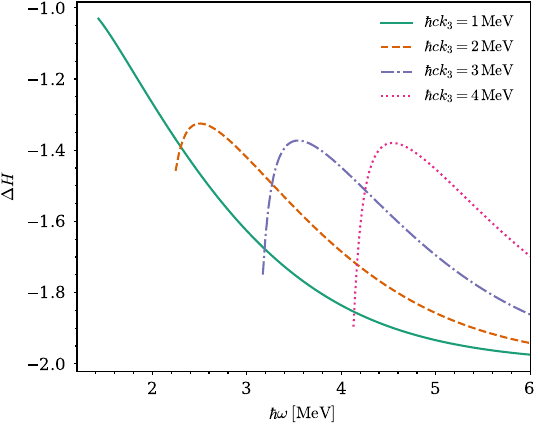}

 \caption{\protect\justifying
Net pair-creation phase-space factor $\Delta H$
in the LLL
as a function of the photon energy\cite{Beloborodov2013,Daugherty1983,Harding2006,TATAROGLU2025} $\hbar\omega$
for different values of the longitudinal momentum. Results are shown for a magnetic field\cite{Harding2006,Turolla2015,Kaspi2017} $B = 7\cdot10^{14}\,\mathrm{G}$,
the chemical potential\cite{Svensson1984} $\mu \simeq mc^2$
and temperature\cite{Svensson1984,Potekhin2015} $T \simeq m c^2$.
Increasing $\hbar ck_3$ shifts the kinematic threshold for
pair creation to higher energies,
reflecting the one-dimensional nature in the LLL with the constraint
$\hbar\omega_{\rm thr}^{(n=0)} \leq \hbar\omega <\hbar\omega_{\rm thr}^{(n=1)}$ for $\hbar ck_3$ fixed.
}
\label{deltaH/w-diferente k3}
 	\end{figure}

 	\begin{figure}[h]
 		\centering
 		\includegraphics[width=8.5cm]{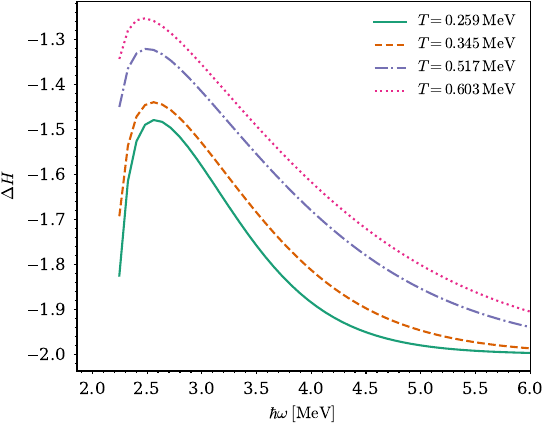}
 \caption{\protect\justifying
Net pair-creation phase-space factor $\Delta H$
in the LLL as a function of the photon energy \cite{Beloborodov2013,Daugherty1983,Harding2006}
$\hbar\omega \leq\, 6\, \mathrm{MeV} < \hbar\omega_{\rm thr}^{(n=1)}$ for different temperatures ($T$ in energy units).
Results are shown for a magnetic field\cite{Harding2006,Turolla2015,Kaspi2017} $B = 7\cdot10^{14}\,\mathrm{G}$,
$\hbar ck_3 = 2\,\mathrm{MeV}$,
and chemical potential\cite{Svensson1984} $\mu \simeq mc^2$.
The temperature values correspond to relativistic pair plasmas relevant in early neutron star environments\cite{Svensson1984,Potekhin2015}.
}\label{deltaH/w-diferente temperatura}
 	\end{figure}

Let us now  discuss the term $n^p(\varepsilon_s)+n^e(\varepsilon_s)-1$ that appears in (\ref{escalar-s}). It is responsible for the behaviour of $\triangle H$ as a function of the radiation field energy. The creation-annihilation balance equation can be written as\cite{Nishida2021}:
 	
 	\begin{equation}\label{ecuacion-balance-creacion aniquilacion}
 		\Gamma(\varepsilon_s)=n^p(\varepsilon_s)n^e(\varepsilon_s)-[1-n^p(\varepsilon_s)][1-n^e(\varepsilon_s)],
 	\end{equation}
 	
 \noindent where the term $[1-n^p(\varepsilon_s)][1-n^e(\varepsilon_s)]$ is the probability that vacant states exist for a fixed $n$, while the $n^e(\varepsilon_s)n^p(\varepsilon_s)$ is associated with the annihilation process. The latter can occur if there are pairs present. Notice that  $\Gamma(\varepsilon_s)=n^p(\varepsilon_s)+n^e(\varepsilon_s)-1$. It should be emphasized, however, that the inequality $n^p(\varepsilon_s)+n^e(\varepsilon_s)<1$ does not follow automatically from $T\neq 0$; it depends on the chemical potential $\mu$. In a regime with moderate values of $\mu$ and $T$ (as in our numerical estimates), the sum is indeed less than unity, and Pauli blocking reduces the pair creation rate\cite{Prakapenia2023}. For sufficiently large $\mu$ (i.e., in a dense medium), one may have $n_p(\epsilon_s) + n_e(\epsilon_s) \geq 1$, which would completely suppress pair creation or even render annihilation dominant. 
 The net pair-creation phase-space factor $\Delta H$, defined from the absorptive part of the polarization tensor, is negative when pair creation dominates over annihilation (e.g., $\Delta H=-2$ in vacuum). The sign convention is set by the expression (\ref{creation-particles}); negative values thus indicate a net pair creation rate, while positive values would correspond to annihilation dominance.
 As can be seen in Fig. \ref{deltaH/w-diferente-n}\renewcommand{\thefootnote}{$\dagger$}\footnotemark\footnotetext{In numerical calculations we restore $\hbar$ and $c$.}\renewcommand{\thefootnote}{\arabic{footnote}}, we have a reduction of the vacuum contribution, that is, $\triangle H$ is less negative. This  reduction is greater when the Landau level is lower. The pair creation threshold for $n=3$,
$\omega_{\rm thr}^{(n=3)}=\sqrt{k_3^2+4m^2+12|e|B}$, exceeds the photon energies considered in our numerical analysis. The $n=3$ channel remains kinematically closed, which further suppresses higher-Landau contributions.

The choice of physical parameters used in the numerical analysis is motivated by conditions expected in strongly magnetized astrophysical environments. Magnetic fields in the range $10^{14}-10^{15}\,\mathrm{G}$ are characteristic of magnetars~\cite{Harding2006,Turolla2015,Beloborodov2012,Kaspi2017}. Photon energies in the MeV range arise naturally in pair cascades and high-energy emission processes in magnetospheres\cite{Beloborodov2013,Daugherty1983,Harding2006,TATAROGLU2025}. Temperatures of order $T \sim m$ correspond to relativistic pair plasmas that may occur in early neutron star evolution\cite{Potekhin2015,Svensson1984}. The longitudinal momentum scale corresponds to relativistic particle propagation along the magnetic field. In strongly magnetized astrophysical environments, such as magnetar magnetospheres, pulsars, and relativistic outflows, electron-positron pairs are typically produced with Lorentz factors\cite{Beloborodov2013} $\gamma \gg 1$. This justifies the representative choice $k_3 \sim\mathrm{MeV}$ adopted in this work.

Notice also, from Fig. \ref{deltaH/w-diferente-n}, that $\Delta H(\omega)$ decreases with $\omega$ and remains finite for  $\omega \gg T, \mu$. In this case, it is valid for $n<n_{B,\omega,k_3}=3$, where $n_{B,\omega,k_3}$ is fixed by $B=7\cdot10^{14}\,\mathrm{G}$, and $k_3=2\, \mathrm{MeV}$ within the considered energy range. The above can be obtained by analyzing (\ref{axial anomaly-pair creation}) for a general value of $n_{B,\omega,k_3}$ in the high-energy limit $\omega \gg \{\sqrt{2eB}, m, k_3\}$. In this limit, kinematics imposes that the Landau number $n$ satisfies $n <\omega^2/(2eB)$. Hence, the larger $\omega$ is compared to $\sqrt{2eB}$, the larger the maximum $n$ that can contribute, and the series is naturally truncated without arbitrary cutoff. We have then

\begin{equation}
k_{\mu} j_{\mu}^A(\omega) \simeq - \frac{i e^{3}}{2\pi} \sum_{n=0}^{n_{B}}\alpha_n \frac{\varepsilon^{2}_{n,0}\,\Delta H(\omega)}{\omega^2}\hspace{1mm}(\textbf{E}^{(2)}\cdot \textbf{B}),
\end{equation}

\noindent where
\begin{equation}
	\Delta H(\omega)\simeq 2\left(  \frac{1}{1+\exp{\frac{\omega/2-\mu}{T}}}+ \frac{1}{1+\exp{\frac{\omega/2+\mu}{T}}}\right)-2.
\end{equation}

We can see that $\Delta H(\omega)/\omega^2$ approaches zero for $\omega \gg T,\mu$, which implies that a transition between chiral-active and 
non-chiral-active regimes could occurs.

Fig. \ref{deltaH/w-diferente k3} illustrates the interplay between longitudinal kinematics and Fermi-Dirac statistics in the pair-creation process induced by longitudinal photons in the LLL. For each value of the longitudinal momentum $k_3$, the pair production is only allowed in the region where energy satisfies the constraint $\omega_{\rm thr}^{(n=0)}< \omega <\omega_{\rm thr}^{(n=1)}$, which sets a $k_3$-dependent energy threshold. The $\Delta H(\omega)$ develops a stationary maximum determined by the condition $d\Delta H/d\omega=0$, which reflects a balance between the explicit $\omega$ dependence of the occupation numbers and their implicit dependence through the kinematic relation $\varepsilon_s(\omega)$. If $k_3\sim 2m$, the Pauli blocking is stronger and the associated peak is the highest in Fig. \ref{deltaH/w-diferente k3}. In an extended intermediate-energy region, variations of $k_3$ merely shift the position of this maximum along the energy axis without appreciably affecting its height, which is fixed solely by Fermi-Dirac statistics and the thermodynamic parameters of the medium. For sufficiently large $k_3$, however, the energy at the stationary point increases and eventually exceeds the thermal scale, leading to an exponential suppression of occupation numbers. As a consequence, the peak height gradually decreases, and the vacuum limit $\Delta H\to -2$ is recovered, indicating that the medium becomes ``transparent'' to pair creation. Here ``transparent'' means that medium effects encoded in the occupation numbers become negligible, and the pair creation rate approaches its vacuum value.

This behavior demonstrates that while the magnetic-field-induced dimensional reduction controls the kinematic accessibility of pair creation, the magnitude of medium-induced suppression is universally governed by quantum statistics.  Fig. \ref{deltaH/w-diferente temperatura} shows how the energy gap is narrower at lower temperatures in the LLL. Thermal effects further modulate this process. This figure highlights how thermal effects modulate the microscopic pair creation process that drives the chiral magnetic effect in a magnetized medium. Here, for the magnetic fields used in our numerical analysis and photon energies $\omega\lesssim 6 \, \mathrm{MeV}$, the $n=1$ channel remains kinematically closed, ensuring the dominance of the LLL.

Our results underscore that chiral transport is not only governed by magnetic field strength and longitudinal kinematics, but is also sensitively dependent on the thermodynamic state of the plasma.
 The present paper aims to show that a chiral current along $\mathbf{B}$ can indeed arise from the pair creation process induced by longitudinal photons.
  This could act as a microscopic source of chiral imbalance and arises purely from the interplay between magnetic-field-induced spatial and chiral symmetry breaking, without requiring either an initial electric charge asymmetry or a chirality imbalance, typically imposed as inputs. The chiral asymmetry  modulates the separation of electric charges along $\textbf{B}$, involving the decay of longitudinal \emph{axial} photons in electron-positron pairs.

\subsection{Regime near the magnetized QED vacuum}
\label{sec:vacuum}

If we consider only the contribution of the LLL in the region
$z_{1} < -4m^{2}$, from (\ref{axial anomaly-pair creation}) we have:

\begin{align}\label{axial anomaly-LLL}
k_{\mu} j_{\mu}^A = - \frac{i e^{3}}{2\pi} \frac{m^{2} \Delta H(\varepsilon_s)}{\sqrt{z_1(z_1+4m^{2})}} (\textbf{E}^{(2)} \cdot \textbf{B}),
\end{align}

\noindent where

\begin{equation}\label{energia-creacio}
\varepsilon_{s} = \frac{\omega}{2} - \frac{|k_{3}|}{2} \sqrt{1 - \frac{4m^2}{|z_{1}|}}.
\end{equation}

In a low-temperature regime near the   magnetized QED vacuum, we have that $\Delta H\simeq -2$. In this limit, we can write

\begin{equation}\label{axial anomaly-limit vacuum}
 k_{\mu} j_{\mu}^A \simeq \frac{i e^{3}}{\pi} \frac{m^{2}}{\sqrt{z_1(z_1+4m^{2})}} (\textbf{E}^{(2)} \cdot \textbf{B}).
\end{equation}

The equation (\ref{axial anomaly-limit vacuum}) suggests that, unlike thermal or density-induced effects, strong magnetic fields can enable the generation of chiral imbalance through photon-induced pair creation in an appropriate low-temperature regime.

The vanishing of (\ref{axial anomaly-limit vacuum}) in the massless limit is specific to the LLL contribution. When summing over  Landau levels $n > 0$, the complete expression  does not vanish in the limit $m \to 0$. The vanishing of (\ref{axial anomaly-limit vacuum}) is therefore an artifact of the LLL truncation. For $n \geq 1$, the terms $2eBn$ remain finite even in the massless limit, ensuring a non-vanishing contribution. Vacuum effects in strong magnetic fields including all Landau levels can be seen in Ref.\cite{Hattori2013}. This behavior connects with  theoretical developments in massless QED in strong magnetic fields~\cite{Kharzeev2016}, where it has been shown that the CME persists for massless fermions.

Equation (\ref{axial anomaly-limit vacuum}) suggests the existence of multiple chiral configurations that can be accessed by varying the parameter $z_1$ of the longitudinal photons. Therefore, this serves as a kinematic control parameter in the magnetized system, where the invariant $\mathbf{E}^{(2)} \cdot \mathbf{B}$ acts as a source for the axial charge. Theoretical work on axion electrodynamics in condensed matter systems~\cite{Li2010, Nandi2018} has observed a similar behavior. Our result provides a fundamental QED basis for the chiral current generation associated with pair creation in the low-temperature regime near the magnetized quantum vacuum. Here, we have configurations parameterized by the chiral charge density, and an electric field aligned with $\mathbf{B}$ may induce transitions between those. This picture is further supported by recent studies of the chiral magnetic effect~\cite{Kharzeev2016}.

\subsection{Longitudinal negative magnetoresistance}
\label{sec:conductivity}

The conductivity associated with the pseudovector mode is proportional to the corresponding eigenvalue of the photon self-energy tensor in the medium. In general, the structure of this current is given in terms of the scattering and pair creation of electrons and positrons resulting from the decay of the longitudinal photons. It is worth noting that while the analysis of Ref.\cite{Baier1996} addressed the general process of photon decay in a magnetized plasma, it did not specifically consider the longitudinal mode, whose unique properties and associated chiral transport are our central interest.

We start from a general expression for the real conductivity, which can be expressed explicitly in terms of the imaginary part of $\Pi_{\mu\nu}$ as \cite{Shabad1,Hugo2,Hugo3,Shabad}:

\begin{equation}\label{conductividad-OP}
  \sigma_{ij}=\frac{Im[\Pi_{ij}]}{\omega}.
\end{equation}

The contribution to the current density  $j_{i}$ in (\ref{corriente-tensor de conductividad}) from conductivity can then be written in the general form $j_{i}=\sigma_{ij}E_{j}+(E\times S)_{i}$, where $\sigma_{ij}=Im[\Pi_{ij}^{S}]/\omega$ and
$S_{i}=\frac{1}{2}\epsilon^{ijk}\sigma_{jk}^{H}$ is a pseudovector
associated with $\sigma_{jk}^{H}=Im[\Pi_{ij}^{A}]/\omega$. Here
$\epsilon^{ijk}$ is the third rank antisymmetric unit tensor, and
$\Pi_{ij}^{S},\Pi_{ij}^{A}$ are the symmetric and antisymmetric
parts of $\Pi_{\mu\nu}$. The
first term corresponds to the
Ohm current and the second is the Hall current  \cite{Lidice}. The Hall and Faraday effects occur for the $C$-non-symmetric case; as different from the CME, which occurs in both the $C$-symmetric and $C$-non-symmetric cases \cite{Acosta2016}. Here we will only work
with the Ohm current term associated with the longitudinally  polarized
mode.

Let us consider the specific case of current along $\textbf{B}$ associated with the longitudinal mode, which is chiral non-symmetric. It can be expressed in the form:

\begin{equation}\label{corriente-j}
  j_{3}=\sigma_{33}^{5}\,E_{3},
\end{equation}

\noindent where $\sigma_{33}^{5}$ is the chiral conductivity. It will be calculated taking into account only the pair creation process. From (\ref{conductividad-OP}), and taking into account the previous relation $\Pi_{33}=s(k_{4}^{2}/z_{1})$, we have:

\begin{equation}\label{conductividad-escalar-s}
  \sigma_{33}^{5}=-\frac{\omega \,Im[s]}{z_{1}}.
\end{equation}

From (\ref{Imag-s}), and \eqref{conductividad-escalar-s}, we derive a general expression for $\sigma_{33}^{5}$  associated with a chiral effect in the region $z_1<-4\varepsilon^2_{0,n}$:

\begin{eqnarray}\label{sigma}
	\sigma^{5}_{33}=\frac{e^3 B \omega}{2\pi z_1}\sum_{n=0}^{\infty}\alpha_n \frac{\varepsilon^{2}_{n,0}\, \Delta H(\varepsilon_s)}{\sqrt{z_1(z_1+4\varepsilon^{2}_{n,0})}}.
\end{eqnarray}

\begin{figure}[h]
 		\centering
 		\includegraphics[width=8.5cm]{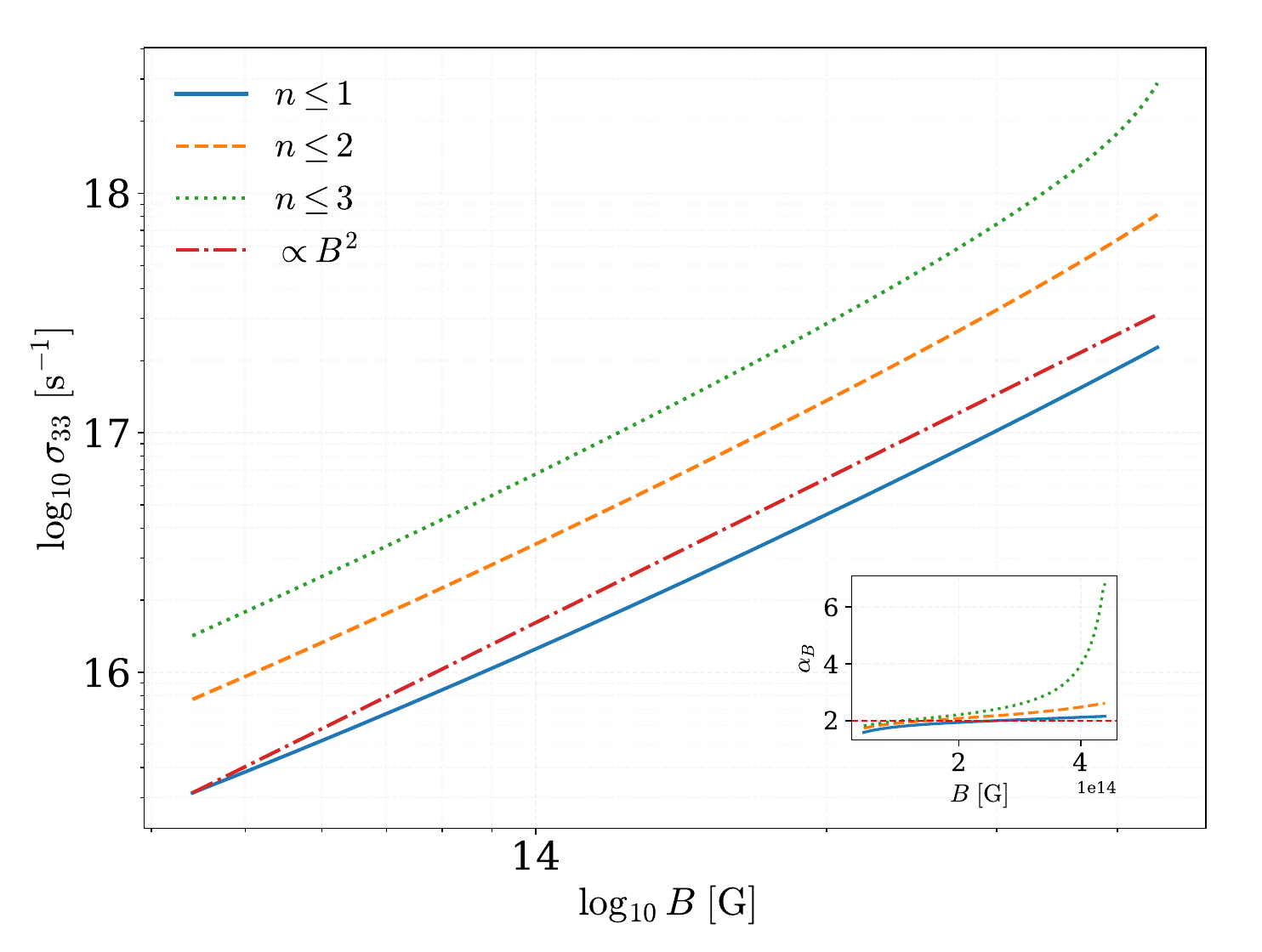}
 \caption{\protect\justifying Longitudinal magnetoconductivity $\sigma_{33}^5$ as a function of the magnetic field $B$ (logarithmic scale).
The different curves correspond to partial sums over Landau levels, including contributions up to $n=0$, $n=1$, $n=2$, and $n=3$.
The inset shows the local scaling exponent $\alpha_B = d\ln\sigma_{33}^5/d\ln B$, highlighting its approach to $\alpha_B \simeq 2$ in the intermediate regime and its deviations outside this range. The parameters used are chemical potential\cite{Svensson1984} $\mu \simeq mc^2$, longitudinal momentum\cite{Beloborodov2013} $ \hbar ck_{3}=2\,\mathrm{MeV}$, photon energy\cite{Beloborodov2013,Daugherty1983,Harding2006,TATAROGLU2025} $\hbar\omega=8.5\,\mathrm{MeV}$, and temperature\cite{Svensson1984,Potekhin2015} $T\simeq mc^{2}$. The explored magnetic field range is consistent with magnetar-strength fields \cite{Harding2006,Turolla2015,Beloborodov2012,Kaspi2017}.
}\label{sigma/B}
 	\end{figure}

The magnetic-field dependence of the longitudinal conductivity is shown in Fig. \ref{sigma/B}. This increases monotonically with the magnetic field strength, indicating a strong enhancement of transport along the direction of the field. This behavior can be understood from the structure of the fermion spectrum in a strong magnetic field. The Landau level spacing grows as $\sqrt{2eB}$, which progressively suppresses the occupation of higher Landau levels (in Fig. \ref{sigma/B} the channels remain kinematically closed for $n>3$). As the magnetic field increases, the fermion dynamics becomes effectively one-dimensional and dominated by the LLL. In this regime the transverse motion is frozen and charged particles propagate primarily along the direction of the magnetic field.

The longitudinal electric field induces pair creation processes that generate charge carriers moving preferentially along $\textbf{B}$. As a consequence, the longitudinal current along field $\mathbf{B}$  increases and the effective resistance decreases, leading to a behavior analogous to negative longitudinal magnetoresistance, which  is caused by the chiral imbalance discussed previously. Although  Fig. \ref{sigma/B} suggests  a near-quadratic dependence on the magnetic field, $\sigma_{33}^5 \propto B^2$ for the lowest Landau levels, this behavior must be interpreted with caution. The scaling is neither exact nor universal, but rather emerges as an \emph{effective} behavior within a restricted range of magnetic field strengths. In this window, the combined effect of dimensional reduction and the approximately linear growth of the Landau degeneracy with $B$ leads to a near-quadratic scaling. However, outside a certain magnetic-field regime associated with the specific parameters used in  Fig. \ref{sigma/B}, clear deviations from $\alpha_B \simeq 2$ are observed. At lower magnetic fields, multiple Landau levels contribute with comparable weight, and the conductivity reflects a more intricate superposition of channels, resulting in a weaker and non-universal scaling. At higher fields, the approach to kinematic thresholds of higher Landau levels (notably the contribution up to $n=3$ in  Fig. \ref{sigma/B}) induces localized distortions in the slope, leading to departures from the quadratic behavior.

Therefore, the apparent $\sigma_{33}^5 \propto B^2$ law should be understood as a \emph{transient scaling regime} rather than a fundamental property of the system. Its validity is controlled by the interplay between the effect of dimensional reduction of field $\textbf{B}$, longitudinal photon kinematics, and the phase-space restrictions imposed by pair creation. This observation highlights that longitudinal negative magnetoresistance associated with chiral effects in magnetized QED plasmas is governed by a richer structure than a simple power-law dependence, and that care must be taken when extracting scaling exponents from finite parameter ranges.

In particular, the breakdown of the quadratic scaling near Landau-level thresholds reinforces the interpretation that transport coefficients are sensitive to the discrete spectrum, and cannot be fully captured by continuum or semiclassical approximations. Notably, the expression (\ref{sigma}) also reveals that the conductivity remains finite as $\mu \rightarrow 0$, demonstrating that the chiral current is $C$-symmetry independent. The effect emerges purely from the combination of spatial symmetry breaking induced by $\mathbf{B}$ and the chiral imbalance associated with the pair creation process. It provides a microscopic link between chiral effects and macroscopic transport coefficients, thereby contributing to the understanding of the chiral magnetic effects with applications in astrophysical scenarios and laboratory plasmas. The conductivity values shown in Fig. \ref{sigma/B} expressed in units of $\mathrm{s}^{-1}$, are consistent with theoretical predictions for relativistic QED plasmas in strong magnetic fields in astrophysical scenarios \cite{Harutyunyan2016,Harutyunyan2024}.

\section{CONCLUSIONS}
\label{sec:conclusions}

In this work we  have proposed a microscopic QED mechanism for the chiral current generation that does not rely on an external chiral chemical potential. Instead, the absorption of longitudinal photons in a strongly magnetized relativistic plasma generates a dynamical chiral imbalance via real electron-positron pair creation. This process is encoded in the imaginary part of the one-loop photon polarization tensor and yields a nonzero divergence of the axial current proportional to \(\mathbf{E}^{(2)}\cdot\mathbf{B}\), while preserving vector current conservation. The resulting longitudinal conductivity \(\sigma_{33}^{5}\) remains finite even for \(\mu = 0\), confirming that the effect is independent of \(C\)-symmetry and arises purely from the interplay of magnetic-field-induced dimensional reduction and the chiral imbalance.

Pauli blocking, controlled by temperature and density, regulates the pair-creation phase space and can completely suppress the chiral current in extreme conditions. Our analysis reveals that the longitudinal conductivity exhibits an approximately quadratic scaling with the magnetic field only within a restricted intermediate regime. This behavior emerges from the dominance of the lowest Landau levels and the kinematic suppression of higher channels. However, clear deviations from this scaling appear outside this window, particularly in the vicinity of higher Landau-level thresholds, where abrupt changes in the slope can be observed. These features reflect the discrete structure of the Landau spectrum and indicate that the magnetoconductivity is highly sensitive to deviations from the expected power-law scaling with the magnetic field, which appear when higher Landau channels become kinematically accessible or when thermal effects strongly modify the occupation numbers.

A particularly insightful limit is analyzed in Section~\ref{sec:vacuum}, where we approach the magnetized QED vacuum at low temperature. In this regime the pair-creation phase-space factor \(\Delta H\) tends to its vacuum value \(-2\), and the
relation for the axial current simplifies. Crucially, while a truncation to the lowest Landau level would yield a vanishing contribution for massless fermions, a summation that includes Landau levels beyond the lowest restores a finite chiral imbalance, underscoring the necessity of including such excited levels in the massless limit. Moreover, the longitudinal parameter \(z_1\) acts as a kinematic control parameter that selects among multiple chiral configurations for each Landau level. This near-vacuum regime suggests that strong magnetic fields 
can enable the generation of chiral imbalance through photon-
induced pair creation in an appropriate low-temperature
regime.

An intriguing feature emerges at high frequencies, as discussed in Section~\ref{sec:mechanism1}. When \(\omega\) becomes much larger than all other scales \{\(T\), \(\mu\), \(k_3\), \(m\), \(\sqrt{2eB}\)\}, the pair-creation phase-space factor \(\Delta H\) remains finite (tending to \(-2\)), yet the axial current divergence scales as \(\Delta H/\omega^2\) and therefore vanishes. This signals a transition from a chiral-active to a non-chiral-active regime: although pairs are still created, their net chiral imbalance becomes ineffective at driving a macroscopic current. This behavior highlights the crucial role of the photon frequency in controlling the chiral transport, and it suggests that high-energy photons, even if abundant, may not contribute to the chiral magnetic current in strongly magnetized plasmas.

More broadly, the mechanism presented here demonstrates that the chiral current can emerge from dissipative quantum processes without any preexisting chirality imbalance. This offers a possible perspective on a broader picture linking the quantum chiral effect to macroscopic transport. The results may have astrophysical implications, as well as relevance for laboratory plasmas and for the quark‑gluon plasma produced in heavy‑ion collisions, where transient magnetic fields are present\cite{Li2026}.


\appendix

\section{}

\subsection{Electromagnetic modes for propagation along \textbf{B}}
\label{sec8}

From the structure of $\Pi_{\mu\nu}$ in a magnetized quantum relativistic system\cite{Hugo2,Hugo3},
in the case of nonvanishing temperature $T$ and chemical
potential $\mu$, we can find the polarization properties of three electromagnetic eigenmodes propagating in the system \cite{Hugo2,Shabad1}. Under those conditions the polarization tensor may be expanded in
 terms of six independent transverse tensors \cite{Hugo3}

 \begin{equation}\label{diagonalize form of the polarization tensor}
   \Pi_{\mu\nu}=\overset{6}{\underset{n=1}{\sum}}\pi^{(i)}\Psi_{\mu\nu}^{(i)}.
 \end{equation}

 As is shown in Ref.
  \cite{Hugo2}, symmetric properties in quantum statistics reduce the number of the basic tensors from an initial set of $9$ to a final set of $6$. The basic tensors are:

\begin{eqnarray}\label{tens-basicos-S}
  \Psi_{\mu\nu}^{(1)} &=& k^{2}T_{\mu\nu},\quad\Psi_{\mu\nu}^{(2)}=(Fk)_{\mu}(Fk)_{\nu}\nonumber\\
  \Psi_{\mu\nu}^{(3)} &=& -k^{2}T_{\mu}^{\lambda} F_{\lambda\eta}^{2}T^{\eta}_{\nu},\quad\Psi_{\mu\nu}^{(4)}=R_{\mu}R_{\nu},
\end{eqnarray}

\noindent with $T_{\mu\nu}=(g_{\mu\nu}-k_{\mu}k_{\nu}/k^{2})$, and $R_{\mu}=(u_{\mu}-k_{\mu}(uk)/k^{2})$; where $u_{\mu}$ is the four-velocity of the medium, and $F_{\mu\nu}$ is the electromagnetic tensor. We express tensor quantities in Minkowski metric, in the form $g_{\mu\nu}=(1,1,1,-1)$, understanding the order $\mu,\nu=1,2,3,0$ as it was agreed after  (\ref{ecuacion/SD}). The tensors (\ref{tens-basicos-S}) are symmetric in the indexes $\mu$, $\nu$ while the following ones are antisymmetric

\begin{eqnarray}
\Psi_{\mu\nu}^{(5)}&=&(uk)[k_{\mu}(Fk)_{\nu}-k_{\nu}(Fk)_{\mu}+k^{2}F_{\mu\nu}]\nonumber\\
\Psi_{\mu\nu}^{(6)}&=&u_{\mu}(Fk)_{\nu}-u_{\nu}(Fk)_{\mu}+(uk)F_{\mu\nu}.
\end{eqnarray}\label{tens-basicos-A}

We introduce a set of orthonormal vectors which are the
  eigenvectors of $\Pi_{\mu\nu}$ in the limit $(\mu,T)\rightarrow(0,0)$

\begin{eqnarray}\label{vect-orton}
c_{\mu}^{(1)}&=& R_{1}(F^{2}k)_{\mu}k^{2}-k_{\mu}(kF^{2}k),\quad c_{\mu}^{(2)}=R_{2}(F^{*}k)_{\mu}\nonumber\\
c_{\mu}^{(3)}&=& R_{3}(Fk)_{\mu},\quad c_{\mu}^{(4)}= R_{4}k_{\mu},
\end{eqnarray}

\noindent where $R_{i}\, (i=1,2,3,4)$ are normalization parameters, and $F^{*}_{\mu\nu}$ is the dual of the electromagnetic tensor
$F_{\mu\nu}$. Using these vectors we can obtain the scalars:

\begin{eqnarray}\label{escalares}
  p&=& c^{(1)\mu}\Pi_{\mu}^{\nu}c^{(1)}_{\nu},\quad s=c^{(2)\mu}\Pi_{\mu}^{\nu}c^{(2)}_{\nu}\\
  t&=& c^{(3)\mu}\Pi_{\mu}^{\nu}c^{(3)}_{\nu},\quad r=c^{(3)\mu}\Pi_{\mu}^{\nu}c^{(1)}_{\nu}
\end{eqnarray}

and the pseudoscalars:

\begin{subequations}\label{psudoescalares}
\begin{eqnarray}
  q&=&c^{(2)\mu}\Pi_{\mu}^{\nu}c^{(1)}_{\nu},\quad v=c^{(2)\mu}\Pi_{\mu}^{\nu}c^{(3)}_{\nu}
\end{eqnarray}
\end{subequations}

From (\ref{diagonalize form of the polarization tensor}) we can find the polarization properties\cite{Hugo2,Shabad1} of three electromagnetic eigenmodes of $\Pi_{\mu\nu}$. In the case of propagation along the magnetic field \textbf{B} in a magnetized medium, the eigenmodes of $\Pi_{\mu\nu}$ are:

\begin{eqnarray}\label{eigenmodes}
b^{(2)}_{\mu}&=&ac^{(2)}_{\mu}\hspace{1mm}e^{i(k_{3}x_{3}-\omega t)}\nonumber\\
b^{(1,3)}_{\mu}&=&b(c^{(1)}_{\mu}\pm ic^{(3)}_{\mu})\hspace{1mm}e^{i(\textbf{k}_{\bot}\cdot\textbf{r}_{\bot}-\omega t)},
\end{eqnarray}

\noindent with eigenvalues $\eta^{(2)}=s$ and $\eta^{(1,3)}=t\pm \sqrt{-r^2}$ respectively. Here $\textbf{r}_{\bot}$ is the coordinate vector in the plane-$(x,y)$, and $a,b$ are parameters, which have dimensions of vector potential. The electric and magnetic fields associated with these modes are obtained from the equations:

\begin{equation}\label{ecuaciones-campo electrico-magnetico}
  \textbf{E}^{(i)}=-\frac{\partial\textbf{b}^{(i)}}{\partial
x_{0}}-\frac{\partial\ b^{(i)}_{0}}{\partial\textbf{x}},\hspace{3mm} \textbf{H}^{(i)}=\nabla\times \textbf{b}^{(i)},
\end{equation}

\noindent with $(i=1,2,3)$. For the case of
$C$-symmetric $(\mu=0)$, the mode $b_{\mu}^{(3)}$ is a transverse
plane polarized wave, whose electric unit vector is
$\textbf{E}_{u}^{(3)}=\textbf{e}_{\perp}\times \textbf{e}_{3}$ ,
orthogonal to the plane $(\textbf{B}, \textbf{k})$, where we define $\textbf{e}_{\perp}=\textbf{k}_{\perp}/k_{\bot}$ and
$\textbf{e}_{3}=\textbf{k}_{3}/k_{3}$ as the transverse and
parallel unit vectors respectively. The mode $b_{\mu}^{(2)}$ is pure electric and
longitudinal with $\textbf{E}_{u}^{(2)}=\textbf{e}_{B}$ (we recall that $\textbf{e}_{B}=\textbf{B}/B$ is a pseudovector), whereas
$b_{\mu}^{(1)}$ is transverse $E_{u}^{(1)}=\textbf{e}_{\perp}$.  In this $C$-symmetric case  $\eta^{(1)}=\eta^{(3)}$ ,
and the circular polarization unit vectors $(\textbf{E}_{u}^{(1)}\pm
i\textbf{E}_{u}^{(3)})/\sqrt{2}$ are common eigenvectors of $\Pi_{ij}$
and of the rotation generator matrix $A^{3ij}$ .

In the case of $C$-non-symmetric $(\mu\neq 0)$. The second mode
$b_{\mu}^{(2)}$ is the same pure longitudinal wave as in the $C$-symmetric case. The
transverse modes $b_{\mu}^{(1,3)}$ describe circularly polarized
waves in the plane orthogonal to $\textbf{B}$  having different
eigenvalues, typical of Faraday effect\cite{Lidice}.

\subsection{Calculation of $Im[s]$}
\label{sec9}

The denominator $D$ of the integral for $s$ ((\ref{escalar-s}), which have singularities due to $D$) is given by:

\begin{equation}\label{Denominador-D}
  D=4z_{1}p_{3}(p_{3}+k_{3})+z^{2}_{1}-4 \omega^{2}\varepsilon^{2}_{n,0},
\end{equation}

\noindent where $z_{1}=k_{3}^{2}-\omega^{2}$ and $\varepsilon_{n,0}^{2}=m^{2}+2|e|nB$, with $n=2n_{l}+1-sgn(e) s_{3}$. It can be written in a form symmetric under the exchange\cite{Hugo3} $\varepsilon_{q}\leftrightarrow \varepsilon_{q^{\prime}}$, $\omega\leftrightarrow-\omega$ 

\begin{multline}\label{denomonador-DI}
 D^{-1}\hspace{-0.1cm}=\hspace{-0.1cm} \frac{1}{8\varepsilon_{q^{\prime}}\varepsilon_{q}\omega} \hspace{-0.1cm}(\frac{1}{\varepsilon_{q^{\prime}}-\varepsilon_{q}-\omega+i\epsilon}
  -\frac{1}{\varepsilon_{q^{\prime}}-\varepsilon_{q}+\omega+i\epsilon} \\
 -\frac{1}{\varepsilon_{q^{\prime}}+\varepsilon_{q}-\omega+i\epsilon}+
\frac{1}{\varepsilon_{q^{\prime}}+\varepsilon_{q}+\omega+i\epsilon}),
\end{multline}

\noindent where $\varepsilon_{q^{\prime}}=\sqrt{(p_{3}+k_{3})^2+m^2+2|e|nB}$ and $\varepsilon_{q}=\sqrt{p_{3}^{2}+m^2+2|e|nB}$, with $q=(n,p_{3})$. The first pair of singularities  are related to excitation of particles to higher energies and the second two are connected to the pair creation. We have added an infinitesimal positive imaginary part $i\epsilon$ to $\omega$, and by using the relation:

\begin{equation}\label{delta}
   \frac{1}{s-\omega-i\epsilon}=P\frac{1}{s-\omega}+i\pi\delta(s-\omega),
\end{equation}

\noindent where $P$ corresponds to the principal value in the expression, we get for the imaginary  part\cite{Hugo3} of  $D^{-1}$ 

\begin{multline}\label{ImD}
  Im D^{-1}\hspace{-0.1cm}=\hspace{-0.1cm} \pm \frac{\pi}{8\varepsilon_q\varepsilon_{q^{\prime}}\omega}  \hspace{-0.1cm}[\delta(\varepsilon_{q^{\prime}}-\varepsilon_q\mp \omega)+\delta(\varepsilon_{q^{\prime}}-\varepsilon_q\pm \omega) \\
 -\delta(\varepsilon_{q^{\prime}}+\varepsilon_q\mp \omega)],
\end{multline}

\noindent where the $\pm$ signs apply respectively to $\omega\gtrless0$. We can use now  (\ref{ImD}) to obtain the imaginary part of the escalar $s$ (see (\ref{escalar-s}))
according to the relation:

\begin{equation}
\int_{-\infty}^{\infty}dp_3f(p_3)\delta(g(p_3))=\sum_m\frac{f(p_3^m)}{\mid g^{\prime}(p_3^m)\mid}\label{formula},
\end{equation}

\noindent where $p_3^m$, with $m=(1,2)$ are the roots of $g(p_3)=0$. It may be easily shown that while $p_3$ runs within $(-\infty<p_{3}<\infty)$, the denominator of the expression (\ref{escalar-s}) may vanish\cite{Hugo3} only for real $z_{1}$. Thus, the integral in (\ref{escalar-s}) represents an analytic function in the $z_{1}$ plane except possible singularities located somewhere on the real axis, which corresponds to the absorption region ( $Im[\Pi_{33}]$ is responsible for the absorption process for the longitudinal mode), where

\begin{equation}\label{momentum}
  p_{3}^{(1,2)}=\frac{-k_{3}z_{1}\pm \omega\Lambda}{2z_{1}},
\end{equation}

\noindent are the roots\cite{Hugo3} of the denominator in (\ref{escalar-s})  and $\Lambda=\sqrt{z_{1}(z_{1}+4\varepsilon^{2}_{n,0})}$. In our case $g(p_3)=\omega\pm(\varepsilon_{q^{\prime}}\pm \varepsilon_{q})$, thus:

\begin{equation}\label{balance energia/momentum}
  \omega=\varepsilon_{q^{\prime}}\pm \varepsilon_{q},\hspace{2mm}k_{3}=p^{\prime}_{3}\pm p_{3},
\end{equation}

\noindent and the corresponding values of the energies are given by:

\begin{equation}\label{energia-ecxitacion}
  \varepsilon_{r}=\frac{-\omega z_{1}+|k_{3}|\Lambda}{2z_{1}},
\end{equation}
\begin{equation}\label{energia-cracion}
  \varepsilon_{s}=\frac{\omega z_{1}+|k_{3}|\Lambda}{2z_{1}},
\end{equation}

\noindent where $r,s=(n,\omega,k_{3})$. The $\pm$ signs in (\ref{balance energia/momentum}) correspond to the pair creation $(\varepsilon_{s})$ and excitation cases $( \varepsilon_{r})$ respectively. By substituting these expressions, it is easy to obtain:

\begin{equation}
\mid \frac{d}{dp_3}(g(p_3))\mid=\frac{\Lambda}{2\varepsilon_q^m\varepsilon_{q^{\prime}}^m}
\label{resultado}.
\end{equation}

In the evaluation of the integral (\ref{escalar-s}) containing the second delta (\ref{ImD}), the following exchange is made $p_{3}+k_{3}\leftrightarrow -p_{3}$, $n^{\prime}\leftrightarrow n$.

\end{document}